\documentclass[aps,preprint]{revtex4}%
\usepackage{amsfonts}
\usepackage{amsmath}
\usepackage{amssymb}
\usepackage{graphicx}%
\begin{document}

\begin{center}
\textbf{Salient features in locomotor evolutionary adaptations of
proboscideans revealed via the differential scaling of limb long bones}

.

By Valery B. Kokshenev and Per Christiansen

.

Submitted to the Journal of the royal Society Interface 03 June 2009

.
\end{center}

\textbf{Abstract.} The standard differential scaling of proportions in limb
long bones (length against circumference) is applied to a phylogenetically
wide sample of the Proboscidea, Elephantidae and the Asian (\textit{Elephas
maximus}) and African elephant (\textit{Loxodonta africana}). In order to
investigate allometric patterns in proboscideans and terrestrial mammals with
parasagittal limb kinematics, the computed slopes (slenderness exponents) are
compared with published values for mammals and studied within a framework of
theoretical models of long bone scaling under gravity and muscle forces. Limb
bone allometry in \textit{E. maximus} and the Elephantidae are congruent with
adaptation to bending and/or torsion induced by muscular forces during fast
locomotion, as in other mammals, whereas limb bones in \textit{L. africana}
appear adapted for coping with the compressive forces of gravity.
Consequently, hindlimb bones are expected to be more compliant than forelimb
bones in accordance with in vivo studies on elephant locomotory kinetics and
kinematics, and the resultant negative limb compliance gradient in extinct and
extant elephants, which contrasts to other mammals, suggests an important
locomotory constraint preventing achievement of a full-body aerial phase
during locomotion. Differences in ecology may be responsible for the subtle
differences observed between African and Asian elephant locomotion, and the
more pronounced differences in allometric and mechanical patterns established
in this study.


\section{Introduction}

Differential scaling of the proportions of the limb long bones in terrestrial
mammals has been studied by many researchers (most often bone length $L$ and
circumference $C$ or diameter $D$) with the aim to establish correlations
between design and posture of mammalian limbs coping with support of mass and
locomotion in the gravitational field. Many attempts to formulate generally
applicable allometric power laws ($L\varpropto M^{l}$ and $C=\pi D\varpropto
M^{d}$, where $M$ is body mass) have been the subject of a long standing
debate and controversy (e.g., McMahon 1973, 1975a, b; Alexander 1977;
Alexander et al. 1979a; Biewener 1983, 2005; Economos 1983; Bertram \&
Biewener 1990; Christiansen 1999a, b, 2002, 2007; Kokshenev, 2003, 2007;
Kokshenev et al. 2003). Among the proposed theoretical frameworks, the three
most common of which are: the \emph{geometric} (or isometric) \emph{similarity
model} (GSM, with $l_{0}=d_{0}=1/3$); the \emph{elastic similarity model}
(ESM, with $l_{0}=1/4,d_{0}=3/8$); and the static \emph{stress similarity
model} (SSM, with $l_{0}=1/5,d_{0}=2/5$). Subsequently, these similarity
models have typically been explored in analysis of allometric power laws using
measured bone lengths and diameters resulted in allometric exponents ($l$ and
$d$), when body masses are available. In cases where body masses are unknown,
a \emph{slenderness exponent} $\lambda$ can be computed (via $L\varpropto
C^{\lambda}$) and compared with that\textrm{ }predicted as $\lambda_{0}%
=l_{0}/d_{0}=1$, $2/3$, or $1/2$ by the GSM, ESM, and SSM, respectively
(McMahon 1975a).\qquad

Originally proposed to explain animal design to experience similar elastic
forces and stresses under gravity, the corresponding ESM (or "buckling" model)
and SSM (McMahon 1973, 1975a, b) were found not to apply to terrestrial
mammals as a group, neither overall body proportions (Economos 1983; Silva
1998), nor allometric scaling of long bones (Alexander et al. 1979a; Biewener
1983, 2005; Economos 1983; Christiansen 1999a, b; Kokshenev et al. 2003).
Instead, body proportions and long bones were found closer to isometry, i.e.,
to GSM with $\lambda_{0}=1$, and in large mammals and in large mammals they
both become progressively more robust with increments in body size. Testing
McMahon's models, many studies indicated that muscle forces, showing
size-dependent fluctuations among terrestrial mammals, are highly important
(e.g., Alexander 1985, Biewener 1983, 1989, 1990, 2005). Direct evidences of
muscle forces affecting proportions of the allometric scaling of the major
long bones in Artiodactyla were provided by Selker and Carter (1989). It was
analytically shown that the failure of any one predicted power law was shown
not caused by the failure of underlying elastic force patterns but due to
McMahon's simplifications for evolutionary adaptive properties for maintaining
similar skeletal functional stresses apart of muscles and under the dominating
influence of gravity (Kokshenev et al. 2003). Then, using the overall-bone
slenderness exponent $\lambda_{\exp}^{(mam)}=0.80\pm0.02$, resulting from
re-analysis of bone proportions in a wide-ranging sample of terrestrial
mammals from Christiansen (1999a), also not matching any of McMahon's original
predictions, the dominating role of muscle forces in long-bone scaling was
demonstrated from first physical principles (Kokshenev 2003).

Naturally, bone shape is a hereditary property, but bone is a phenotypically
plastic tissue, capable of reacting powerfully to its mechanical environment.
In the growing fetus, bone shape, size and position is initially determined by
the early cartilaginous anlagen during embryonic skeletogenesis, which are
subsequently gradually replaced by endochondral ossification during ontogeny
(Favier \& Doll\'{e} 1997; Currey 2003; Provet \& Schipani 2005). However,
bone shape is heavily influenced by a mechanical response to the environment
during ontogeny and throughout an animal's life. It has been demonstrated that
strain rate and magnitude, surrounding tissue formation, and fetal muscle
contractions are prerequisites for normal bone formation during ontogeny
(Rodriguez et al. 1992; Mosley et al. 1997; Mosley \& Lanyon 1998; Lamb et al.
2003). In post-natal and adult mammals, bone is capable of reacting to changes
in mechanical stresses enforced by physical activity and muscle mechanics with
rapid alterations of size and shape (Biewener 1983, 1989, 1990; Carrano \&
Biewener 1999; Curry 2003; Firth et al. 2005; Warden et al. 2005; Franklyn et
al. 2008), and hereditary properties determined by the genome appear primarily
responsible for bone patterning during fetal ontogeny and is less relevant for
bone size and shape in the adult animal (Mariani \& Martin 2003). These
factors are seemingly beyond macroscopic elastic theories for formulating
allometric power laws for bone scaling. Moreover, the concept of uniform
elastic similarity also seems to be inconsistent with a diversity of
functional local elastic forces and stresses, which are not constant in bones
during locomotion and therefore are not likely to be reflected in scaling
analysis of external bone dimensions (Doube et al. 2009). Nevertheless, the
authors hopefully believe that scaling predictions arising from the
macroscopic spatial continuous mechanics applied to bone tissue under broad
spectrum of loading conditions (Kokshenev et al. 2003), and therefore
reflecting most general trends in proportional limb bone adaptations to
environmental conditions, can be reliably verified at least by the overall
limb and bone allometric data.

The above bone scaling studies and other statistical and experimental studies
(e.g., Biewener et al. 1983a, b, 1989; Rubin \& Lanyon 1984; Selker \& Carter,
1989; Streicher \& Muller 1992; Carrano \& Biewener 1999) have stimulated
formulation of novel theoretical concepts in light of dynamic bone strain
similarity (Rubin \& Lanyon, 1984) or mechanical (strain and stress)
similarity (Kokshenev 2007). Basic conceptions of McMahon's (1973, 1975a)
elastic similarity hypothesis have also been reconsidered (Kokshenev et. al.,
2003, Kokshenev 2003). The framework of bone scaling is, by default, limited
to long bones approximated by cylinders with $L\gg D$ justified by the ratio
$L/D\backsim10$, at least for mammalian humerus, radius, ulna, femur, and
tibia. Moreover, the justification for application of the elastic theory
patterns established for arbitrary loaded long solid cylinders (Kokshenev
2007) is generally based on the assumption that the long bones play the
primary role in body support. Consequently, the positive allometry of long
bone structure in relation to body mass observed in regression analysis may be
expected to be better understood by biomechanical adaptation of bones to
maximal external loads emerging during fast locomotion and studied via
bone-reaction elastic forces and stresses, whereas non-mechanical ontogeny
effects of limb bone adaptation associated with a small Prange's index are
relatively small (Kokshenev 2007).

In the present study we analyze the surprisingly varied differential scaling
of the limb long bones in a taxonomically narrow clade of mammals, the extant
proboscideans (Proboscidea, Elephantidae); the Asian elephant (\textit{Elephas
maximus}), and African savannah elephant (\textit{Loxodonta africana}). These
gigantic land mammals have a more upright limb posture, notably much more
upright propodials and different locomotor mechanics from other terrestrial
mammals in that fast locomotion is ambling with no suspended phase in the
stride, but with duty factors $\beta>0.5$ (Gambaryan 1974; Alexander et al.
1979b; Hutchinson et al. 2003, 2006). We compare theoretical predictions with
the data from a phylogenetically wide sample of extinct proboscideans from
Christiansen (2007) completed here by the Elephantidae family, as well as
allometry results from scaling studies of running mammals with parasagittal
limb kinematics, in the hope of establishing generic allometric patterns
distinguishing limb postures characteristic of high-power locomotion in
proboscideans and mammals.

Modern elephants are characteristic in having very long limb bones for their
body size (Christiansen 2002), and a very upright, though not strictly
columnar (Ren et al. 2007), limb posture, in which the two propodial bones
(femur and in particular humerus) are kept at a distinctly greater angle
compared to the ground than is the case in other large, quadrupedal mammals.
The steeply inclined propodials imply that during standing and at low speeds,
the primary forces affecting the limb bones will be axially compressive.
However, with increments in speeds, joint flexion increases during the support
phase, and during the recovery phase, joint flexion can be high (Ren et al.
2007). Although there is no difference in joint flexion between juvenile and
adults or between Asian and African elephants, the propodial bones are still
markedly more inclined compared to horizontal even during the fast locomotion
than is in the case for other quadrupedal running mammals. During the support
phase in locomotion, the ankle, unlike the more mobile wrist, displays
spring-like mechanical properties, reminiscent of, albeit less than in
quadrupedal running mammals, matching the more compliant hind limbs during
locomotion. This also is consistent with the tendons of the hind-foot scaling
with positive allometry during ontogeny, whereas those of the forelimb scale
with negative allometry, and thus become progressively more gracile (Miller et
al. 2008), thereby supporting the observation that the hindlimbs are more
compliant with bouncing kinematics than the stiffer, vaulting forelimbs,
during fast locomotion. In view of the progress in application of bone scaling
models (Kokshenev 2003, 2007), a general problem arises whether kinematic
evidences on the mechanical influence on bone ontogeny can be independently
revealed by the differential limb bone scaling?

\section{Materials and Methods}

\subsection{Theoretical background}

A theoretical analysis of non-critical elastic forces emerging in long bones
of adult mammals resulted in mode-independent relationships for bone scaling
exponents
\begin{equation}
d=\frac{1}{3}+b\text{ and }l=\frac{1}{3}-b\text{, with }\lambda(b)=\frac{l}%
{d}=\frac{1-3b}{1+3b} \label{beta}%
\end{equation}
discussed in Eq. (8) in Kokshenev 2007. Here $b$ is Prange's index scaling
bone mass to body mass. Since scaling index $b$ is consistently small for
mammals (see e.g. table 1 in Kokshenev 2003), equation (\ref{beta}) matches
previous observations mentioned in Introduction of the closeness of mammalian
bone allometry to isometry. Accordingly, small deviations from the
force-isotopic GSM are described via the directly observable model-independent
index $b$. The differential scaling data for mammals (Christiansen 1999a,
1999b) supports the physically justified inequalities $d>1/3>l$ (Kokshenev
2007) providing the constraint $\lambda(b)<1$, resulting in $b<1/6$ following
from equation (\ref{beta}). The empirically established constraint $b_{\exp
}^{(mam)}=0.04\pm0.01$, for overall-averaged long bones in mammalian limbs
(see table 1 in Kokshenev 2007), results in the \emph{model-independent}
pattern, namely
\begin{equation}
d_{pre}^{(mam)}=0.37\pm0.01\text{ and }l_{pre}^{(mam)}=0.29\pm0.01\text{, or
}\lambda_{pre}^{(mam)}=0.785\pm0.025 \label{lambda}%
\end{equation}
predicted for mammals capable of true running with a fully suspended aerial
phase in the stride $[1]$.\footnotetext[1]{It seems to be interesting to
compare the proposed empirical exponent $\lambda_{\operatorname{mod}}%
^{(mam)}=0.785$, substituting McMahon's $\lambda_{0}=1/2$, with the
theoretical estimate $\lambda_{\operatorname{mod}}=7/9$ ($\thickapprox0.778$)
obtained on the basis of a pattern of non-axial elastic forces in bone
resisting functionally relevant limb muscles (Kokshenev 2008). The bone-muscle
scaling theory will be discussed elsewhere.}

\qquad As for McMahon's models discussed in Introduction and revised by
Kokshenev (2003, 2007) and Kokshenev et al. (2003), they can be broadly
interpreted as follows. If both gravitational and muscular competitive forces
driven by structural bone adaptation to complex (axial and non-axial)
compression were equally important in bone interspecific allometry, the
observed statistically overall-bone slenderness exponent $\lambda_{\exp}$ is
expected to be nearly isometric, i.e., close to maximum $\lambda_{0}=1$,
following from the GSM. If, however, gravitational forces were dominating,
bone proportions could be expected to become optimized for exploitation of
long bone stiffness during modes of fast locomotion in mammals with near
parasagittal limb kinematics, and would result in $\lambda_{0}=2/3$, as
predicted by the ESM, as predicted by the ESM, also known as the buckling
model. Although the ESM was originally introduced for the states of elastic
instability (McMahon 1973) or thermodynamic instability (Kokshenev et al.
2003), the domain of resulting scaling relations extends far below critical
amplitudes of forces (and critical stresses and strains), which should still
exclude non-axial elastic forces (Kokshenev 2007). In contrast, the adaptation
to faster forms of locomotion via more compliant long bones subjected to the
non-axial compression could be observed via the exponent $\lambda_{0}=1/2$,
predicted by the SSM, treated as a "bending-torsion" model (Kokshenev 2007).
As an example of theoretical rationalization underlying these two distinct
patterns of bipedal locomotion in the sagittal plane, a transition from
stiff-limbged slow-walking to the compliant-limbged fast walking following by
running was illuminated in terms of a dynamic instability of the trajectory of
center of gravity in humans (Kokshenev 2004).

When muscle forces play a dominating role in the formation of bone
proportions, the appropriate modified model for bone evolution (hereafter
termed SSMM) predicts $\lambda_{pre}^{(bend)}=0.80\pm0.03$, as derived from
both allometric mammalian limb bone and muscle data (see equation (18) in
Kokshenev 2003). Since this prediction is congruent with the one made in
equation (\ref{lambda}), we infer that mammalian long bones are designed to
resist peak bending and/or torsional bone compressions produced by muscles
during fast running modes, that was initially established via bending
functional bone stress (Rubin \& Lanyon 1982) and then explained analytically
(see figure 1 in Kokshenev 2007). For mammals as group, we therefore use the
SSMM estimate $\lambda_{\operatorname{mod}}^{(mam)}=0.785$ predicted
semi-empirically in equation (\ref{lambda}) and theoretically in $[1]$.

\subsection{Materials and analysis}

We used previously published data on external limb long bone dimensions from
$19$ species and $217$ specimens of proboscideans (Christiansen 2007), which
was supplemented by new data collected for the purpose of this study. We
compared this to published data from $79$ and $98$ species of running mammals
from Christiansen 1999a and 1999b, respectively. We conducted regression
analysis on $Log_{10}$ transformed external limb long bone articular lengths
and diaphysial diameters using the standard \emph{least squares} (LS) and
\emph{reduced major axis} (RMA) methods. All species with multiple specimens
were averaged prior to analysis. The significances of the regression
parameters were evaluated by computing the correlation coefficient, the
standard error of the estimate, and the F-statistic of the regressions, and
the $95\%$ confidence intervals for the regression intercepts and slopes (see
online electronic supplementary material). The regression analysis included
the entire Proboscidea; the Elephantidae (\textit{Elephas} sp.,
\textit{Loxodonta africana}, and \textit{Mammuthus} sp.). We computed separate
regression analyses for the extant Asian elephant (\textit{Elephas maximus})
and African savannah elephant (\textit{Loxodonta africana}), since these are,
by default, the only taxa for which locomotory information exists; we included
no data from the forest elephant (\textit{Loxodonta cyclotis}) within the
African elephant, because this taxon most likely constitutes a separate
species (Barriel et al. 1999).

\section{Results}

Searching for generic morphometric patterns in long bones via external
dimensions, to which the similarity models (McMahon 1973, 1975a; Kokshenev
2003, 2007; Kokshenev et al. 2003) are broadly addressed, we study the
slenderness bone exponent $\lambda$ as directly observed by the slopes in
plots $Log_{10}L$ vs $Log_{10}C$. In table 1, the results of bone-size
regression analysis in species-averaged specimens of proboscideans are
compared with those for mammals.

.%

\begin{tabular}
[c]{|l|rll||lll|rll|}\hline
Species & Ele & phant & idae & Pro & bosci & deans & \multicolumn{1}{||r}{Ma}
& mmals & \\\hline\hline
Limb bones & \multicolumn{1}{|c|}{$N$} & \multicolumn{1}{|c|}{$\lambda$} &
\multicolumn{1}{|c||}{$r$} & $N$ & \multicolumn{1}{|c|}{$\lambda$} &
\multicolumn{1}{|c||}{$r$} & \multicolumn{1}{|c}{$N$} &
\multicolumn{1}{|c|}{$\lambda$} & \multicolumn{1}{|c|}{$r$}\\\hline
Humerus & \multicolumn{1}{|l|}{7} & \multicolumn{1}{|l}{0.912} &
\multicolumn{1}{|l||}{0.990} & 16 & \multicolumn{1}{|l}{\textit{1.134}} &
\multicolumn{1}{|l||}{0.831} & \multicolumn{1}{|l}{189} &
\multicolumn{1}{|l}{0.7631} & \multicolumn{1}{|l|}{0.9738}\\\hline
Radius & \multicolumn{1}{|l|}{6} & \multicolumn{1}{|l}{0.813} &
\multicolumn{1}{|l||}{0.853} & 10 & \multicolumn{1}{|l}{\textit{1.078}} &
\multicolumn{1}{|l||}{0.878} & \multicolumn{1}{|l}{189} &
\multicolumn{1}{|l}{0.7530} & \multicolumn{1}{|l|}{0.9957}\\\hline
Ulna & \multicolumn{1}{|l|}{6} & \multicolumn{1}{|l}{0.727} &
\multicolumn{1}{|l||}{0.888} & 14 & \multicolumn{1}{|l}{0.929} &
\multicolumn{1}{|l||}{0.866} & \multicolumn{1}{|l}{189} &
\multicolumn{1}{|l}{0.849$^{\ast}$} & \multicolumn{1}{|l|}{0.9600}\\\hline
Femur & \multicolumn{1}{|l|}{7} & \multicolumn{1}{|l}{0.747} &
\multicolumn{1}{|l||}{0.966} & 14 & \multicolumn{1}{|l}{0.802} &
\multicolumn{1}{|l||}{0.816} & \multicolumn{1}{|l}{189} &
\multicolumn{1}{|l}{0.8431} & \multicolumn{1}{|l|}{0.9763}\\\hline
Tibia & \multicolumn{1}{|l|}{6} & \multicolumn{1}{|l}{0.751} &
\multicolumn{1}{|l||}{0.925} & 11 & \multicolumn{1}{|l}{0.772} &
\multicolumn{1}{|l||}{0.857} & \multicolumn{1}{|l}{188} &
\multicolumn{1}{|l}{0.7641} & \multicolumn{1}{|l|}{0.9499}\\\hline
Limb bone, LS & \multicolumn{1}{|l|}{6} & \multicolumn{1}{|l}{\textbf{0.790}}
& \multicolumn{1}{|l||}{0.924} & 13 & \multicolumn{1}{|l}{\textbf{0.943}} &
\multicolumn{1}{|l||}{0.850} & \multicolumn{1}{|l}{189} &
\multicolumn{1}{|l}{\textbf{0.795}} & \multicolumn{1}{|l|}{0.971}\\\hline
Limb bone, RMA & \multicolumn{1}{|l|}{6} & \multicolumn{1}{|l}{\textbf{0.856}}
& \multicolumn{1}{|l||}{0.924} & 13 & \multicolumn{1}{|l}{\textbf{1.165}} &
\multicolumn{1}{|l||}{0.850} & \multicolumn{1}{|l}{189} &
\multicolumn{1}{|l}{\textbf{0.778}} & \multicolumn{1}{|l|}{0.971}\\\hline
\end{tabular}

\textbf{Table 1}. The statistical data on the slenderness of individual and
effective limb bones in animals. The data for Elephantidae and Proboscidea are
shown on the basis of regression data provided in online electronic
supplementary material, extending table 2 in Christiansen 2007, and these for
mammals are taken from table 2 in Christiansen 1999b. The mean
\emph{slenderness exponents} $\lambda=dLog_{10}L/dLog_{10}C$ presented by the
slopes ($\lambda$) exemplified in figure 1 are observed in $N$\textrm{
}\emph{species} through the LS regression of with the \emph{correlation
coefficient} $r$. The limb bone LS characterization corresponding to the
overall-bone mean data is introduced by the standard mean over all 5 bones.
The RMA data in the last row are shown only for the resulting limb bone data.
The bold numbers are the data used below in figures. The italic numbers
indicate the slope data contrasting to mammalian data with $\lambda<1$ (see
also discussion in section 2.1 in the Methods). $^{\ast}$) The data estimated
with the help of ratio $l/d$ for the ulna allometric exponents taken from
table 2 in Christiansen 1999a.

.

In figure 1 and table 2, we analyze modern elephants.

.

\textbf{Place figure 1}

\textbf{.}%

\begin{tabular}
[c]{|l|rll|lllrl|}\hline
Species & \textit{E.} & \textit{maxi} & \textit{mus} &
\multicolumn{1}{||l}{\textit{Loxo}} & \textit{donta} &  & \textit{afri} &
\textit{cana}\\\hline\hline
Limb bones & \multicolumn{1}{|c|}{$n$} & \multicolumn{1}{|c|}{$\lambda$} &
\multicolumn{1}{|c||}{$r$} & \multicolumn{1}{|c}{$n$} &
\multicolumn{1}{|c|}{$\lambda$} & \multicolumn{1}{|c}{$r$} &
\multicolumn{1}{|c|}{$p_{\min}$} & \multicolumn{1}{|c|}{$p_{\max}$}\\\hline
Humerus & \multicolumn{1}{|c}{22} & \multicolumn{1}{|l}{0.754} &
\multicolumn{1}{|l||}{0.985} & \multicolumn{1}{|c}{14} &
\multicolumn{1}{|l}{0.616} & \multicolumn{1}{|l}{0.978} &
\multicolumn{1}{|l|}{0.01} & \multicolumn{1}{|l|}{0.02}\\\hline
Radius & \multicolumn{1}{|c}{19} & \multicolumn{1}{|l}{\textit{1.014}} &
\multicolumn{1}{|l||}{0.987} & \multicolumn{1}{|c}{8} &
\multicolumn{1}{|l}{0.675} & \multicolumn{1}{|l}{0.994} &
\multicolumn{1}{|l|}{---} & \multicolumn{1}{|l|}{0.001}\\\hline
Ulna & \multicolumn{1}{|c}{20} & \multicolumn{1}{|l}{0.818} &
\multicolumn{1}{|l||}{0.981} & \multicolumn{1}{|c}{11} &
\multicolumn{1}{|l}{0.644} & \multicolumn{1}{|l}{0.979} &
\multicolumn{1}{|l|}{0.01} & \multicolumn{1}{|l|}{0.02}\\\hline
Femur & \multicolumn{1}{|c}{25} & \multicolumn{1}{|l}{0.758} &
\multicolumn{1}{|l||}{0.972} & \multicolumn{1}{|c}{13} &
\multicolumn{1}{|l}{0.618} & \multicolumn{1}{|l}{0.986} &
\multicolumn{1}{|l|}{0.01} & \multicolumn{1}{|l|}{0.02}\\\hline
Tibia & \multicolumn{1}{|c}{20} & \multicolumn{1}{|l}{0.913} &
\multicolumn{1}{|l||}{0.970} & \multicolumn{1}{|c}{10} &
\multicolumn{1}{|l}{0.688} & \multicolumn{1}{|l}{0.939} &
\multicolumn{1}{|l|}{0.02} & \multicolumn{1}{|l|}{0.05}\\\hline
Forelimb bone & \multicolumn{1}{|c}{20} & \multicolumn{1}{|l}{0.862} &
\multicolumn{1}{|l||}{0.984} & \multicolumn{1}{|c}{11} &
\multicolumn{1}{|l}{0.645} & \multicolumn{1}{|l}{0.984} &
\multicolumn{1}{|l|}{0.01} & \multicolumn{1}{|l|}{0.02}\\\hline
Hindlimb bone & \multicolumn{1}{|c}{23} & \multicolumn{1}{|l}{0.836} &
\multicolumn{1}{|l||}{0.971} & \multicolumn{1}{|c}{12} &
\multicolumn{1}{|l}{0.653} & \multicolumn{1}{|l}{0.963} &
\multicolumn{1}{|l|}{0.015} & \multicolumn{1}{|l|}{0.035}\\\hline
Limb bone, LS & \multicolumn{1}{|c}{21} & \multicolumn{1}{|l}{\textbf{0.851}}
& \multicolumn{1}{|l||}{0.979} & \multicolumn{1}{|c}{11} &
\multicolumn{1}{|l}{\textbf{0.648}} & \multicolumn{1}{|l}{0.975} &
\multicolumn{1}{|r}{0.013} & \multicolumn{1}{|l|}{0.028}\\\hline
Limb bone, RMA & \multicolumn{1}{|c}{21} & \multicolumn{1}{|l}{\textbf{0.869}}
& \multicolumn{1}{|l||}{0.979} & \multicolumn{1}{|c}{11} &
\multicolumn{1}{|l}{\textbf{0.665}} & \multicolumn{1}{|l}{0.975} &
\multicolumn{1}{|r}{0.013} & \multicolumn{1}{|l|}{0.028}\\\hline
\end{tabular}

\textbf{Table 2}. The statistical data on $Log_{10}L$ vs $Log_{10}C$
regression for individual limb bones in \textit{Elephas maximus} and
\textit{Loxodonta africana}. Notations of table 1 are extended by the t-test
comparisons of slopes in $n$ \emph{specimens} shown by $p_{min}<p<p_{max}$
(for details see the electronic supplementary material). Characteristic of a
given group of species \emph{forelimb} and \emph{hindlimb bones} are
introduced through the fore-bone (humerus, radius, and ulna) and hind-bone
(femur and tibia) standard means, respectively. The effective \emph{limb bone}
is determined by the overall (5-bone) standard mean. Other notations are the
same as in table 1.

.

The main results obtained by the LS regression in tables 1 and 2 are displayed
and analyzed in figure 2.

.

\textbf{Place figure 2}

.

In figure 2, the model predictions for bone slenderness exponent are compared
with those of the entire group of proboscideans, Elephantidae, mammals, and
modern elephants. As seen from the data presented by the bone-averaged
exponents (in table 2) resulting from the LS and RMA species-average
statistics (shown by bars), the limb bones of the family Elephantidae are
structurally designed likewise those in mammals. The data for extant
\textit{Elephas maximus} are also quite similar to mammals, whereas the bone
exponents in \textit{Loxodonta africana} are distinctly lower (see also figure
1). This implies that the data for \textit{Elephas maximus} as well as the
family Elephantidae, are better explained by adaptations to peak muscular
forces during locomotion, whereas the limb bones in \textit{Loxodonta
africana} are indicative of adaptation to cope with the forces of gravity
$[2]$.\footnotetext[2]{Our sample of \textit{Elephas maximus} has more large
juveniles included than in \textit{Loxodonta africana}, and all resulting
slopes are generally higher than in \textit{Loxodonta africana} (table 2).
Comparing only adult specimens, therefore ignoring ontogenetic adaptations,
the radius and femur slopes remain significantly lower in \textit{Loxodonta
africana} than \textit{Elephas maximus}, whereas the exponents for humerus,
ulna and tibia become non-significantly different (see online supplementary
material). In \textit{Loxodonta africana}, all exponents for adult specimens
only are similar to the full sample including large juveniles, whereas the
slopes in \textit{Elephas maximus} become significantly higher (femur) and
lower (tibia). Overall, the LS-average slenderness exponent in adult
\textit{Elephas maximus} ($\lambda_{\exp}=0.839$) remains significantly higher
($p<0.05$) than in adult \textit{Loxodonta africana} ($\lambda_{\exp}%
=0.662$).} The Elephantidae and individual species within this family also
have thinner long bones than more primitive proboscideans (Haynes 1991;
Christiansen 2007).

.

\textbf{Place figure 3}

.

In figure 3, we examine the allometry exponents of individual bones, with the
aim to interpret their adaptation to the patterns of peak elastic forces and
stresses. Within the Elephantidae, the similarity in variation of bone
slenderness in those three groups is very evident from the parallel lines, as
shown in figure 3. When the mechanical origin of the model predictions
provided in Methods is taken in consideration, the limb bones in
\textit{Elephas maximus} indicate adaptation to complex muscular and, in part,
gravity stresses.

It is well established in comparative zoology that the humerus of running
parasagittals is loaded differently owing to large muscle attachments and an
inclined angle compared to the epipodials, but in elephants the humerus and
femur are not very steeply inclined. This is broadly congruent with the
observation in figure 3 that the femur, which being almost vertical when the
animal stands motionless and much more inclined in running mammals than in
fast moving elephants, should involve lower bending and torsional moments in
\textit{Elephas maximus} than in other mammals and much less muscular moments
in \textit{Loxodonta africana}. Indeed, as predicted by the ESM beyond
experimental error, the limb bones in \textit{Loxodonta africana} appear to be
adapted for axially compressive stress generated by gravity. Such a
distinction in allometry slopes induced by mechanical adaptation, clearly
distinguishes \textit{Loxodonta africana} from \textit{Elephas maximus},
which, in turn appears to bear a mechanical, if not morphological similarity
to the limb bones of mammals capable of true running. Accordingly, the limbs
of \textit{Elephas maximus} would appear to be more adapted for resisting the
forces from limb muscles, which broadly is more consistent with faster forms
of locomotion.

As seen in figure 3, the radius and femur in modern elephants expose different
loading trends of those in mammals. A consequent distinct mechanical
characterization of the hindlimbs and forelimbs is displayed in figure 4.

\qquad\textbf{Place figure 4}

.

In figure 4, the differently designed forelimb and hindlimb bones are
schematically shown through the hind-fore-limb bone vector, indicating the
existence of a gradient in the limb bone functions, likewise gradients in
muscle functions\textrm{ }in mammals (Biewener et al. 2006) increasing the
stability of running (Daley et al. 2007). One can see that the Elephantidae
contrasts to mammals in limb bone functions.

\section{Discussion}

\subsection{Surveying kinematic empirical data}

Recent studies have provided new insights and have significantly enhanced our
understanding of elephant locomotion. Typically called graviportal, elephants
are unable to run with a suspended phase in the stride, and even during fast
locomotion, one limb is placed firmly on the ground (Gambaryan 1974). New
studies have, however, indicated that fast moving elephants are not merely
walking, but that kinetics and kinematics during fast locomotion differ from
walking, and numerous locomotor parameters are similar to those of running
mammals, for instance offsetting the of phase of forelimb and hindlimb
footfalls; achieving the transient walk-run duty factors $0.5$; and
maintenance of pendular locomotor kinematics, typical of walking gaits, for
the forelimbs, whereas the hind limbs move with a more spring-like (bouncing)
action (Hutchinson et al. 2006; Ren \& Hutchinson 2007). Accordingly, at high
speeds elephant locomotor kinematics are indicative of walking gaits whereas
kinetic analyses indicate running, as in other quadrupedal mammals. The
primary differences are that elephants never achieve a full-body aerial phase
during any form of locomotion, although at speeds of just over $\symbol{126}2$
$ms^{-1}$, the hind limbs begin exhibiting an aerial phase with bouncing
kinetics, implying that the limbs become compliant, whereas the forelimbs
maintain a more straight morphology, consistent with more vaulting mechanical
properties (Ren \& Hutchinson 2007). Another significant difference from
running, quadrupedal mammals, there is no marked change of gait at high
speeds, even at Froude numbers $Fr>3$, a value when other quadrupedal animals
have changed gaits to a bouncing, running gats with a full-body aerial phase
(Alexander 1983, 1989; Alexander \& Jayes 1983).

During progressively faster locomotion, elephants initially increase speeds
primarily by increments in stride frequency, but at high speeds, further speed
increase is facilitated primarily by increments in stride length (Hutchinson
et al. 2006). This slow-to-fast gait transition is similar to walking-trotting
transition in quadrupedal running mammals, where increments in speed is a
function of those in both stride length and frequency, whereas in the case of
running gaits with bouncing limb kinetics and a full-body aerial phase is
characterized primarily by increments in stride length (e.g., Heglund et al.
1974; Pennycuick 1975, Biewener 1983; Alexander 1983, 1989; Alexander \& Jayes
1983). Locomotor kinematic parameters in Asian and African elephants are
broadly similar, but statistically significant differences exist, pertaining
to relative stride lengths, stride frequencies, stance phase, and duty factor
with speed. African elephants have higher duty factors, shorter stride lengths
and higher stride frequencies than Asian elephants (Hutchinson et al. 2006).
Interestingly, large elephants, such as full grown bulls, appear incapable of
reaching the same locomotor intensity as small elephants, and have duty
factors $\beta<0.5$, implying that no limb pair ever exhibits an aerial phase;
they are, in effect, only walking during fast locomotion. Consequently, this
observation may imply that large individuals may exploit the compliance of
bone tissue to a lesser extent.

\subsection{What does the overall-bone statistics tell us?}

The theoretical concepts provided in the Methods section permits an
interpretation of the limb bone allometric data in terms of the adapted
loading patters. Beyond any modeling, the regression analysis illustrated in
figure 1 indicates that the individual limb bones in distinct extant elephants
are likely similar in bone ontogeny but they are evidently different in their
mechanical adaptation as revealed by distinct slopes in size proportions. When
known theoretical models are employed in terms of the overall-bone slenderness
exponent (in figure 2), one can see that the limbs in \textit{Elephas
maximus}, as well as in Elephantidae, show a similarity to the limbs of other
mammals, broadly exploiting muscular forces during efficient locomotion. This
observation agrees with \textit{in vivo} data on the elephant locomotor
kinematics having generally many patterns in common with typical tetrapods
(e.g., Hutchinson at al. 2006). In contrast, our analysis show that
\textit{Loxodonta africana} successfully employs body gravitation and reaction
gravitation forces for efficient walking. More specifically, the analysis in
figure 2 indicates that the overall-bone averaged slenderness allometry
exponent associated with the structure of an effective limb bone (defined in
table 2) can be understood in mammals and the Elephantidae (but not
intraspecifically in \textit{Loxodonta africana}), by its adaptation to peak
muscular forces generated during fast locomotion, whereas the limb bones in
\textit{Loxodonta africana} are indicative of adaptation to cope with the
forces of gravity more successfully exploiting in walking. This finding also
implies that both propodial and epipodial limb bones in mammals are
established to be adapted for peak functional bending and torsional stresses
(figure 3) by the exploiting of bone compliance, contrasting to more stiffer
limbs in Asiatic elephants, as predicted by the ESM (figure 4).

When \textit{Elephas maximus} and \textit{Loxodonta africana} are compared, no
contrasting adaptation of any individual limb bone is revealed in figure 3,
because all bone lines are parallel. This is not the case of lines lying
between mammals as group and elephants. This observation suggests a similarity
between bone joint angles generally established for modern elephants during
fast locomotion by direct observations, e.g., by Ren et al. (2007).
Consequently, bone angles, which are evidently similar in elephants, and
distinct from other mammals, indicate differing loading conditions related to
limb postures in elephants, extant and extinct, from those of other mammals.
More specifically, the epipodial femur, shows in figure 3 its adaptation to
complex compression (bending and/or torsion) during bouncing kinematics of the
hindlimb involved in fast locomotion, thereby exploiting rather the bone
compliance, than the bone stiffness associated with more isometric bone
proportion scaling exposing by the forelimb radius, in \textit{Elephas
maximus} and forelimb humerus in Elephantidae, thereby contrasting to
parasagittal femur and radius. Such mechanical trends are consistent with the
kinematic data (Hutchinson et al. 2006) that the hind limbs of modern
elephants during fast locomotion are broadly more compliant than the fore
limbs. This empirical finding in modern elephants is now generalized over
extinct elephants.

The contrasting postures between mammals as group and elephants can be
understood by the different design of fore limbs and hind limbs revealed in
figure 4 by the opposed directions of the gradient of the limb functions in
locomotor kinematics. As for the small negative gradient in \textit{Loxodonta
africana} with respect to \textit{Elephas maximus}, it can be ignored, because
the typical statistical error (shown by the bars) exceeds the length of the
limb bone vector. In other words, the crossover of the bone lines between the
modern elephants can be referred to a small statistical uncertainty that can
be ignored, providing a qualitative agreement with the overall similarity in
limb postures revealed in figure 3. On the contrary, the observation in figure
4 of the gradient of limb functions for the Elephantidae exceeding the
statistical error may suggest a trend for forelimb bones to be more isometric,
and also that the forelimb bones in \textit{Elephas maximus} are therefore
stiffer than bones in the hind limbs. This also is congruent with the
differences in limb locomotor kinematics (Hutchinson et al. 2006), outlined above.

\subsection{Positive gradient of limb stiffness as major locomotor constraint
in elephants}

Our study of the allometry of individual limb bones reveals different patterns
in limb mechanical adaptation in proboscideans and begs the question how they
correlate with kinematic patterns characteristic of modern elephants? Such a
correlation is expected, since the mean data on duty factor $\beta$ in limbs
of modern elephants (Hutchinson et al. 2006) are underlaid by the overall-bone
(and overall-muscle) limb characterization described here though the limb bone
slenderness exponent $\lambda.$

\qquad Near the walk-run transition in mammals, with the Froude number
$Fr\thickapprox1$, the scaling predictions for the forelimb duty factor
$\beta_{FL}=0.52$ and the hindlimb duty factor $\beta_{HL}=0.53$ (estimated on
the basis of empirical scaling relations by Alexander \& Jayes 1983), provide
negative \emph{limb duty factor gradient} $\Delta\beta$ ($\equiv\beta
_{FL}-\beta_{HL}$) for mammals, contrasting with the positive gradient
$\Delta\beta_{ele}$ for elephants (see table 6 in Hutchinson et al. 2006).
These data can be related to our LS bone exponent $\lambda_{FB}=0.788$ in
mammalian forelimb and $\lambda_{HB}=0.804$ in mammalian hindlimb,
corresponding to the mammalian overall-bone exponent $\lambda_{\exp}%
^{(mam)}=0.795$ (table 1) and the negative \emph{limb bone gradient}
$\Delta\lambda_{\exp}^{(mam)}=-0.016$ (with, $\Delta\lambda\equiv\lambda
_{FL}-\lambda_{HL}$). For elephants, the duty factor gradient $\Delta\beta>0$
also correlates to the bone gradient $\Delta\lambda>0$ (figure 4). Indeed, as
follows from tables 1 and 2, $\Delta\lambda_{\exp}^{(ele)}=0.026$, for
\textit{Elephas maximus}, and $\Delta\lambda_{ele}=0.068$, for Elephantidae.
As for the discrepancy in signs between $\Delta\beta_{\exp}^{(ele)}=0.026$
(table 4 by Hutchinson et al. 2006) and $\Delta\lambda_{\exp}^{(ele)}=-0.008$
for the \textit{Loxodonta africana}, it was referred above to the statistical error.

Employing the similarity in limb functions and their gradients observed
directly (kinematically) in modern elephants and indirectly (allometrically)
via limb bones of extinct and extant elephants, we develop a simple linear
model predicting limb duty factors for Elephantidae, which includes more
primitive groups of proboscideans (see the electronic supplementary material).
The linearization procedure of the mean limb data and their gradients known
for living elephants (shown, respectively, by the dashed lines in figure 5 and
its inset) results in the duty factor predictions for Elephantidae, as
explained in figure 5.

.

\textbf{Place figure 5}

\textbf{.}

The geometrical visualization of locomotory constraints imposed on animal
limbs in a certain locomotion mode can be displayed on a $\lambda$-$\beta$
diagram presented by figure 5. Such a characterization makes a link between
the limb bone and limb bone gradient bone proportions (figure 4) and the limb
and limb gradient kinematics (figure 4 by Hutchinson et al. 2006).

The major limb functional difference in mammals as a group and elephants
indicated by different orientation of the characteristic vectors shown in
figure 5 is due to the difference in signs of the limb stiffness-compliant
gradient transferred between fore and hind limbs during the animal's forward
propulsion of the body. Running mammals, having hindlimbs which are stiffer
than the forelimbs and therefore transfer the positive compliance limb
gradient (or negative stiff limb gradient), but they are able to achieve a
full-body aerial phase during fast locomotion. In contrast, elephants,
transferring negative limb compliance gradient (hindlimb bones are more
compliant than forelimb ones) do not achieve a full-body aerial phase during
any form of locomotion, though are able to sufficiently reduce the positive
stiff limb bone gradient by limb muscles when showing a negative limb duty
factor gradient $\Delta\beta$ during both slow and fast walking gaits (see
figure 4B by Hutchinson et al. 2006). However, it is not enough for changing
of the positive direction of the limb gradient vector in the $\lambda$-$\beta$
diagram, contrasting to mammals (figure 5), since elephants, being naturally
constrained in limb bone proportions, are not able to change sign of the
positive gradient in limb bone slenderness $\Delta\lambda$. Hence, the
preserved excessive positive hindlimb-forelimb stiffness gradient ensured by
the corresponding mass-independent and speed-independent positive limb bone
slenderness, explains the inability of elephants to perform true running with
a full-body aerial phase discussed by Hutchinson et al. 2003. Consequently, in
order to move fast they increase stride frequency, mostly exploiting forelimb
stiffness, instead of compliance, and in order to increase the stride length,
they are forced to use hindlimb bone compliance, instead of stiffness.

\subsection{Asian compared with African elephants}

Being similar in fast locomotion gaits with respect to the lateral sequence
footfall pattern, the Asian, African and most likely extinct elephants (as
predicted figure 5) are found to differ in limb bone and perhaps also muscle
constraints. All having similar projections of the characteristic vector in
the $\lambda$-$\beta$ diagram, the limb bone stress indicated by the bone
slenderness for the African elephant is significantly distinct from that in
other elephants. According to the SSMM, in the Elephantidae and in particular
in \textit{Elephas maximus}, the bone off-axial external muscle forces
generated during fast locomotion, broadly exceeding body weight and causing a
complex bending-torsion elastic bone stress, provide a relatively high level
of limb compliance conducted by the structurally adapted limb long bones. In
contrast, limb bones in \textit{Loxodonta africana} generally adapted for
axial bone compression, are most likely tuned by limb muscles to employ better
gravitation reaction forces, in accord with relatively low bone slenderness,
explained by the ESM.

Having long bones designed to maintain axial stress and avoiding bending and
torsion, African elephants can be expected to exhibit shorter stride lengths
and therefore to use higher stride frequencies than Asian elephants, at
increased locomotor speeds. From an energetic point of view, this implies the
higher energy cost of the Asian elephant locomotion, whereas higher duty
factors characteristic of African elephants (figure 5) indicate less bending
moments about the joints accommodated ground reaction forces. We infer that
Asian elephants, having more compliant limb bones than African elephants, are
broadly able to maintain higher speeds more easily that is statistically
supported by the observation (via $\beta<0.5$) of Asian elephants (figure 4A
by Hutchinson et al. 2006). Even the the limb duty factors in Asian and
African elephants may achieve those in mammals, the negative gradient for the
limb bone compliance limits the maximum stride length and therefore the
maximal running speed with respect to mammals.

It is traditionally believed that mechanical differences must be large to
produce differences in bone morphology (Frost 1990), but more recent studies
have demonstrated that temporal continuous stimulation three orders of
magnitude below the maximal peak forces characteristic of fast locomotion (see
Rubin \& Lanyon 1982) is likely sufficient to produce significant changes in
bone morphology (Rubin et al. 2001). African elephants are typically found in
open environments and routinely undertake long-distance seasonal migrations at
leisurely paces, whereas Asian elephants are mostly found in topologically
more heterogeneous, forested environments and appear to undertake fewer and
shorter, if any, seasonal migrations (Sikes 1971; McKay 1973; Laws et al.
1975; Sukumar 1991, 1992). Potentially, this could imply subtle differences
even in low-force every-day locomotor mechanics imposed by the structure of
the environment. Thus, differences in ecology and migratory activity may
conceivably be responsible for the subtle differences in locomotor mechanics
between African and Asian elephants, as observed by Hutchinson et al. (2006).
In this study, more pronounced differences in allometric and mechanical
patterns are demonstrated for \textit{Elephas maximus}, which appear more
similar to those in other mammals, and \textit{Loxodonta africana}, which is
divergent from both and also from the Elephantidae.

On the other hand, neither allometric nor kinematic or kinetic studies deal
with forces and bone stresses directly. It remains therefore a challenge to
further analyze the reactive-force elastic-stress patterns revealed for limb
bones in extant elephants. Nevertheless, there is another difference between
the two species of extant elephants congruent with our findings. Because the
limb bone pattern for African elephants indicates axial bone stress, not
increasing with body mass (Rubin \& Lanyon 1982, 1984) and therefore would
constitute non-critical stress (Kokshenev 2007), both the mean and maximal
body masses for Asian elephants are expected to be below of those for African
elephants. Indeed, African elephants appear to be larger on average than Asian
elephants; African elephant large bulls routinely weight $5-7$ tons, whereas
$4-5$ tons is more common for Asian elephant bulls (Wood 1976; Shoshani 1991).
Tentative maximal size of African elephant also appears to be distinctly
larger. World recorded bulls are as close to or even exceeding 4 m in standing
shoulder height, and estimated body masses of $10-12$ tons (Wood 1976;
McFarlan 1992), whereas Asian elephants are estimated at $3.3-3.4$ m in
shoulder height and around 8 tons (Pilla 1941; Wood 1976; Blashford-Snell \&
Lenska 1996).  

.

\textbf{Acknowledgments }

One of the authors (V.B.K.) acknowledges financial support by the CNPq.

.

\textbf{Literature}

Alexander, R. McN. 1977 Allometry of the limbs of antelope (Bovidae).
\textit{J. Zool.} \textbf{183}, 125 146.

Alexander; R. McN. 1983 Animal mechanics. Blackwell Scientific Publ., Oxford.
pp. 301.

Alexander, R. McN. 1985 The maximum forces exerted by animals. \textit{J. Exp.
Biol.} \textbf{115}, 211 238.

Alexander, R. McN. 1989 Optimization and gaits in the locomotion of
vertebrates. \textit{Physiol. Rev.} \textbf{69}, 1199 1227.

Alexander, R. McN., Jayes, A. S., Maloiy, G. M. O. \& Wathuta, E. M. 1979a
Allometry of the limb bones of mammals from shrews (\textit{Sorex}) to
elephant (\textit{Loxodonta}). \textit{J. Zool.} \textbf{189}, 305 314.

Alexander, R. McN., Maloiy, G. M. O., Hunter, B., Jayes, A. S. \& Nturibi, J.
1979b Mechanical stresses in fast locomotion of buffalo (Syncerus caffer) and
elephant (\textit{Loxodonta africana}). \textit{J. Zool.} \textbf{189}, 135 144.

Alexander, R. McN. \& Jayes, A. S. 1983 A dynamic similarity hypothesis for
the gaits of quadrupedal mammals. \textit{J. Zool.} \textbf{201}, 135 152.

Barriel, V., Thuet, E. \& Tassy, P. 1999 Molecular phylogeny of Elephantidae.
Extreme divergence of the extant forest African elephant. \textit{Comptes
Rendus de L'Academie des Sciences} \textit{(Ser. III)} \textbf{322}, 447 454.

Bertram, J. E. A. \& Biewener, A. A. 1990 Differential scaling of the long
bones in the terrestrial Carnivora and other mammals. \textit{J. Morphol.}
\textbf{204}, 157 169.

Biewener, A. A. 1983 Allometry of quadrupedal locomotion: the scaling of duty
factor, bone curvature, and limb orientation to body size. \textit{J. Exp.
Biol.} \textbf{105}, 147 171.

Biewener, A. A. 1989 Mammalian terrestrial locomotion and size. Mechanical
design principles define limits. \textit{Bioscience} \textbf{3}, 776 783.

Biewener, A. A. 1990 Biomechanics of mammalian terrestrial locomotion.
\textit{Science} \textbf{250}, 1097 1103.

Biewener, A. A. 2005 Biomechanical consequences of scaling. \textit{J. Exp.
Biol.} \textbf{208}, 1665 1676.

Biewener, A. A., Thomason, J., Goodship, A. \& Lanyon, L. E. 1983a Bone stress
in the horse forelimb during locomotion at different gaits: A comparison of
two experimental methods. \textit{J. Biomech.} \textbf{16}, 565 576.

Biewener, A. A., Thomason, J., \& Lanyon, L. E. 1983b Mechanics of locomotion
and jumping in the forelimb of the horse (\textit{Equus}): in vivo stress
developed in the radius and metacarpus. \textit{J. Zool.} \textbf{201}, 67-82.

Biewener, A. A., Thomason, J., \& Lanyon, L. E. 1988 Mechanics of locomotion
and jumping in the forelimb of the horse (\textit{Equus}): in vivo stress in
the tibia and metatarsus. \textit{J. Zool.} \textbf{214}, 547-565.

Biewener, A. A., Yoo, E. \& McGuigan, M. P. 2006 Is there a proximal-distal
gradient of limb muscle function? \textit{J. Biomech., Suppl.} \textit{1}
\textbf{39}, S358 S358.

Blashford-Snell, J. \& Lenska, R. 1996 \textit{Mammoth hunt: In search for the
giant elephants of Bardia}, London: Harper Collins.

Daley, M. A. , Felix, G. \& Biewener, A. A. 2007 Running stability is enhanced
by a proximo-distal gradient in joint neuromechanical control. \textit{J. Exp.
Biol.} \textbf{210}, 383 394.

Carrano, M.T. \& Biewener, A.A. 1999 Experimental alteration of limb posture
in the chicken (\textit{Gallus gallus}) and its bearing on the use of birds as
analogs for dinosaur locomotion. \textit{J. Morph.} \textbf{240}, 237 249.

Christiansen, P. 1999a Scaling of the limb long bones to body mass in
terrestrial mammals. \textit{J. Morph}. \textbf{239}, 167 190.

Christiansen, P. 1999b Scaling of mammalian long bones: small and large
mammals compared. \textit{J. Zool.} \textbf{247}, 333 348.

Christiansen, P. 2002 Mass allometry of the appendicular skeleton in
terrestrial mammals. \textit{J. Morph}. \textbf{251}, 195 209.

Christiansen P. 2007 Long-bone geometry in columnar-limbed animals: allometry
of the proboscidean appendicular skeleton. \textit{Zool. J. Linn. Soc.}
\textbf{149}, 423 436.

Currey, J.D. 2003 The many adaptations of bone. \textit{J. Biomech.}
\textbf{36}, 1487 1495.

Doube, M., Wiktorowicz, Conroy, A., Christiansen, P., Hutchinson, J.R. \&
Shefelbine, S. 2009 Three-Dimensional Geometric Analysis of Felid Limb Bone
Allometry \textit{PLoS ONE} \textbf{4}, in press (DOI 10.1371/journal.pone.0004742).

Economos, A.C. 1983 Elastic and/or geometric similarity in mammalian design?
\textit{J. Theor. Biol}. \textbf{103}, 167 172.

Favier, B. \& Doll\'{e}, P. 1997 Developmental functions of mammalian Hox
genes. \textit{Mol. Human Repr.} \textbf{3}, 115 131.

Firth, E. C., Rogers, C. W., Doube, M. \& Jopson, N. B. 2005 Muscoskeletal
responses of 2-year old thoroughbred horses to early training. 6. Bone
parameters in the third metacarpal and third metatarsal bones. \textit{New
Zealand Veter. J.} \textbf{53}, 101 112.

Franklyn, M., Oakes, B., Field, B., Wells, P. \& Morgan, D. 2008 Section
modulus in the optimum geometric predictor for stress fractures and medial
tibial stress syndrome in both male and female athletes. \textit{Am. J. Sports
Med.} \textbf{36}, 1179 1189.

Frost, H. 1990 Skeletal structural adaptations to mechanical usage (SATMU): 1.
Redefining Wolff's Law: The bone modeling problem. \textit{Anat. Rec.}
\textbf{26}, 403-413.

Gambaryan, P. P. 1974 How mammals run. Anatomical adaptations. John Wiley \&
Sons, New York.

Haynes, G. 1991 \textit{Mammoth, mastodonts, and elephants. Biology, behavior,
and the fossil record}, Cambridge: Cambridge Univ. Press.

Heglund, N.C., Taylor, C.R. \& McMahon, T.A. 1974 Scaling stride frequency and
gait to animal size: Mice to horses. \textit{Science} \textbf{186}, 1112 1113.

Hutchinson, J. R., Famini, D., Lair, R.. \& Kram, R. 2003 Are fast-moving
elephants really running? \textit{Nature} \textbf{422}, 493 494.

Hutchinson J. R., Schwerda D., Famini D. J., Dale R. H. I., Fischer M. S., \&
Kram R. 2006 The locomotor kinematics of Asian and African elephants: changes
with speed and size. \textit{J. Exp. Biol.} \textbf{209}, 3812 3827.

Kokshenev, V. B. 2003 Observation of mammalian similarity through allometric
scaling laws. \textit{Physica A} \textbf{322}, 491 505.

Kokshenev, V. B., Silva, J. K. L. \& Garcia, G. J. M. 2003 Long-bone allometry
of terrestrial mammals and the geometric-shape and elastic-force constraints
of bone evolution. \textit{J. Theor. Biol.} \textbf{224}, 551 555.

Kokshenev V. B. 2007 New insights into long-bone biomechanics: Are limb safety
factors invariable across mammalian species? \textit{J. Biomech.} \textbf{40},
2911 2918.

Kokshenev V. B. 2008 A force-similarity model of activated muscle is able to
predict primary locomotor functions. \textit{J. Biomech.} \textbf{41}, 912 915.

Lamb, K. J., Lewthwaite, J. C., Lin, J., Simon, D., Kavanagh, E.,
Wheeler-Jones, C. P. D. \& Pitsillides, A. A. 2003 Diverse range of fixed
positional deformities and bone growth restraint provoked by flaccid paralysis
in embryonic chicks. \textit{Intern. J. Exp. Pathol.} \textbf{84}, 191 199.

Laws, R. M., Parker, I. S. C. \& Johnstone, R. C. B. 1975 \textit{Elephants
and their habits. The ecology of elephants in North Bunyoro, Uganda}, Oxford:
Clarendon Press.

Mariani, F.V. \& Martin G. R. 2003 Deciphering skeletal patterning. Clues from
the limb. \textit{Nature} \textbf{423}, 319 325.

McFarlan, D. 1992 \textit{The Guiness book of records}, Enfield, Middlesex:
Guiness Publ. Ltd.

McKay, G. M. 1973 Behaviour and ecology of the Asiatic elephant in
southeastern Ceylon. \textit{Smithsonian Contr. Zool.} \textbf{125}, 1 113.

McMahon, T. A. 1973 Size and shape in biology. \textit{Science} \textbf{179},
1201 1204.

McMahon, T. A. 1975a Using body size to understand the structural design of
animals: quadrupedal locomotion. \textit{J. Appl. Physiol.} 3\textbf{9}, 619 627.

McMahon, T. A. 1975b Allometry and biomechanics: limb bones of adult
ungulates. \textit{Am. Natur.} \textbf{107}, 547 563.

Miller C. E., Basu C., Fritsch G., Hildebrandt T. \& Hutchinson J. R. 2008
Ontogenetic scaling of foot musculoskeletal anatomy in elephants. \textit{J.
R. Soc. Interface} \textbf{5}, 465 475.

Mosley, J. R. \& Lanyon, L. E. 1998 Strain rate as a controlling influence on
adaptive modelling in response to dynamic loading of the ulna in growing male
rats. \textit{Bone} \textit{23}, 313 318.

Mosley, J. R., March, B. M., Lynch, J. \& Lanyon, L. E. 1997 Strain magnitude
related changes in whole bone architecture in growing rats. \textit{Bone}
\textbf{20}, 191 198.

Olmos, M., Casinos, A. \& Cubo, J. 1996 Limb allometry in birds. \textit{Ann.
Sci. Natur. Zool.} \textbf{17}, 39 49.

Pennycuick, C. J. 1975 On the running of the gnu (\textit{Connochaetes
taurinus}) and other animals. \textit{J. Exp. Biol.} \textbf{63}, 775 799.

Pilla, N.G. 1941 On the height and age of an elephant. \textit{J. Bombay Nat.
Hist. Soc.} \textbf{42}, 927-928.

Provot, S. \& Schipani, E. 2005 Molecular mechanisms of endochondral bone
development. \textit{Biochem. Biophys. R. Commun.} \textbf{328}, 658 665.

Ren, L., Butler, M., Miller, C., Paxton, H., Schwerda, D., Fischer, M. S. \&
Hutchinson, J. R. 2007 The movements of limb segments and joints during
locomotion in African and Asian elephants. \textit{J. Exp. Biol.}
\textbf{211}, 2735 2751.

Ren, L. \& Hutchinson, J. R. 2008 The three-dimensional locomotor dynamics of
African (\textit{Loxodonta africana}) and Asian (\textit{Elephas maximus})
elephants reveal a smooth gait transition at moderate speed. \textit{J. Royal
Soc. Interface} \textbf{5}, 195 211.

Rodriguez, J. L., Palacios, J., Ruiz, A., Sanchez, M., Alvarez, I. \& Emiguel,
E. 1992 Morphological changes in long bone development in fetal akinesia
deformation sequence: an experimental study in curarized rat fetuses.
\textit{Teratol.} \textbf{45}, 213 221.

Rubin, C. T. \& Lanyon, L. E. 1982 Limb mechanics as a function of speed and
gait: a study of functional stains in the radius and tibia of horse and dog.
\textit{J. Exp. Biol.} \textbf{101}, 187 211.

Rubin, C. T. \& Lanyon, L. E. 1984 Dynamic strain similarity in vertebrates;
an alternative to allometric limb bone scaling.\textit{ J. Theor. Biol.}
\textbf{107}, 321 327.

Rubin, C. T., Turner, A. S., Bain, S., Mallinckrodt, C. \& McLeod, K. 2001 Low
mechanical signals strengthen long bones. \textit{Nature} \textbf{412}, 603 604.

Selker, F. \textit{\&} Carter, D. R. 1989. Scaling of long bone fracture
strength with animal mass. \textit{J. Biomech.} \textbf{22}, 1175 1183.

Shoshani, J. 1991 Anatomy \& physiology. In \textit{The illustrated
encyclopedia of elephants} (eds. Rogers, G. \& Watkinson, S.), pp. 30-47.
London: Salamander Books.

Sikes, S. K. 1971 \textit{The natural history of the African elephant},
London: Weidenfeld \& Nicholson.

Silva, M. 1998 Allometric scaling of body length: elastic or geometric
similarity in mammalian design. \textit{J. Mammal.} \textbf{79}, 20 32.

Streicher, J. \& M\"{u}ller, G. B. 1992 Natural and experimental reduction of
the avian fibula: Developmental thresholds and evolutionary constraint.
\textit{J. Morph.} \textbf{214}, 269 285.

Sukumar, R. 1991 Ecology. In \textit{The illustrated encyclopedia of
elephants} (eds. Rogers, G. \& Watkinson, S.), pp. 78-101. London: Salamander Books.

Sukumar, R. 1992 \textit{The Asian elephant: ecology and management}.
Cambridge: Cambridge Univ. Press.

Warden, S. J., Hurst, J. A., Sanders, M. S., Turner, C. H. Burr, D. B. \& Li,
J. 2005 Bone adaptation to a mechanical loading program significantly
increases skeletal fatigue resistance. \textit{J. Bone Min. Res.} \textbf{20},
809 816.

Wood, G.L. 1976 \textit{The Guiness book of animal facts and feats}, Enfield,
Middlesex: Guiness Superlatives Ltd.\newpage

\textbf{Figure Captions}

.

Figure 1. Long bone articular lengths against diaphysial least circumferences
in extant elephants. A, humerus; B, ulna; C, femur; D, tibia. Closed squares
are \textit{Elephas maximus}; open squares are \textit{Loxodonta africana}.
Regression coefficients are shown in table 2.

.

Figure 2. A comparison of the predictions by the theory of similarity with the
bone slenderness exponents observed via the regression slopes in different
groups of proboscideans. Notations: open circles are McMahon's predictions for
bones adapted for the influence of gravity ($\lambda_{0}=1,2/3$, and $1/2$ due
to GSM, ESM, and SSM); the closed circle shows mean data $\lambda
_{pred}=0.785$ predicted for the limb bone, which is primarily adapted for
resisting peak muscle forces during locomotion [discussed in equation (2) ].
The bars are the mean LS and RMA data for the limb bone characteristic of
Proboscidea, Elephantidae, and mammals taken from table 1 and these for
\textit{Elephas maximus} and \textit{Loxodonta africana}, taken from table 2.
The error bars show statistical variations between the means of regression data.

.

Figure 3. The observation of trends in limb bone mechanical adaptation in
different species. The bars show maximal variations of the mean exponents of
individual limb bones. The slenderness exponents in humerus (H), radius (R),
ulna (U), femur (F), and tibia (T) are analyzed in view of elastic similarity
models. The notations on the symmetry of bone compression follow from the
models described in Methods. Other notations correspond to those in figure 2.

.

Figure 4. Observation of elastic similarity in the effective forelimb bone
(humerus, radius, and ulna) and hindlimb bone (femur and tibia). The arrows
indicate deviations in the trends of adaptation for forelimb and hindlimb
mechanical functions. The bar shows statistical error. Other notations
correspond to those in fig. 3.

.

Figure 5. Limb bone scaling in mammals and elephants against limb kinematics
in fast walking. The vector positions and magnitudes indicate the slenderness
exponent and duty factor and the vector directions indicate their
forelimb-hindlimb gradients. The dashed vector position predicts the limb duty
factor $\beta_{pre}=0.58$ consistent with $\lambda_{\exp}=0.790$ for
Elephantidae obtained by liner interpolation between the kinematic data for
Asian and African elephants, as shown by the thin dashed line. The
corresponding the model duty factor gradient $\Delta\beta_{\operatorname{mod}%
}=0.043$ is found in the inset, through the linear extrapolation (shown by the
dashed line) of the gradient known for \textit{Elephas maximus} (blue point)
and $\Delta\lambda_{mod}=0.008$ adopted for the \textit{Loxodonta africana}
(green point). Other data are provided above and/or taken from table 4 by
Hutchinson et al. 2006.

\ \ \ \ \ \ \ \ \ 
\end{document}